\documentclass[JJAP]{jjap3}
\usepackage{newtxtext,newtxmath}
\usepackage{braket}
\usepackage{bm}

\title{Elastic constant of dielectric nano-thin films using three-layer resonance studied by picosecond ultrasonics}
\author{Hiroki Fukuda, Akira Nagakubo,\thanks{E-mail: nagakubo@prec.eng.osaka-u.ac.jp} and Hirotsugu Ogi}
\inst{Graduate School of Engineering, Osaka University, Suita, Osaka 565-0871, Japan}
\abst{Elastic constants and sound velocities of nm-order thin films are essential for designing acoustic filters. However, it is difficult to measure them for dielectric thin films. In this study, we use a three-layer structure where a dielectric nano-thin film is sandwiched between thicker metallic films to measure the longitudinal elastic constant of the dielectric film. We propose an efficiency function to estimate the optimal thicknesses of the components. We use Pt/NiO/Pt three-layer films for confirming our proposed method. The determined elastic constant of NiO deposited at room temperature is smaller than the bulk value by $\sim$40\%. However, it approaches the bulk value as the deposition temperature increases. We also reveal that uncertainty of the elastic constant of the Pt film insignificantly affects the accuracy of the determined elastic constant of NiO in this structure.}

\begin{document}
\maketitle

\section{Introduction}
Elastic constants and sound velocities of thin films are essential parameters of surface acoustic wave (SAW) filters\cite{SAWf-LN2009, SAWf-LN2011, SAWf-MultiMode, JJAP-1}, film bulk acoustic resonators (FBAR)\cite{FBAR-JJAP2008, FBAR-JJAP2010_Hashi, FBAR-JJAP2010}, and acoustic mirrors\cite{SAWf-SMR_Ana, SAWf-Murata_gapSMR} for wireless communication. Next-generation communication uses much higher frequency band-pass filters\cite{SAWf-3.5GHzLLSAW, FBAR-JJAP2020}, which are composed of much thinner films. For example, in our previous study, we showed that a 10.9-nm NiO and 7.6-nm Pt multilayer achieved 140-GHz resonance.\cite{USE2020}. It is then important to measure elastic constants and sound velocities of nm-order thin films to precisely control the resonator frequency. Picosecond ultrasonics can directly excite and detect ultrasound with frequencies up to the THz range in metallic and semiconductor thin films.\cite{Laser1, Laser2}. Using this method, we have measured the longitudinal elastic constants of metallic thin films thinner than $\sim$100 nm by observing the through-thickness resonances or pulse echoes, and we reveal that they are usually smaller than those of bulk vales by 5--30\%.\cite{PhononReso2, PU-ML-Nakamura, NA-bW} This softening in thin films indicates that the bulk values should not be adopted in designing the acoustic filters, and the elastic constants of thin films should be measured.

However, it is difficult to measure the elastic constants of dielectric nm-order thin films. Amorphous SiO$_2$ is used as the temperature compensate material\cite{SAWf-TC1998, SAWf-TC2002} because it has positive temperature dependence of sound velocity.\cite{SiO2-TD-v0, SiO2-TD-v2} Acoustic reflectors consist of two materials with high and low acoustic impedance, where dielectric thin films are mainly used as the low-impedance materials.\cite{SMR-JAP, SMR-SiOC} By depositing a $\sim$10 nm metallic film, picosecond ultrasonics can excite ultrasound in dielectric material through the thermal expansion of the metal.\cite{PU-SiO2_atten-Maris} For ``thick'' nm films (>400 nm), we have measured the longitudinal elastic constants and sound velocities of SiO$_2$\cite{PU-Shagawa, NA-SiO2-JAP}, SiON\cite{NA-SiON}, and AlN\cite{NA-JJAP_AlN} films by observing Brillouin oscillation.\cite{BO-Devos_2004, NA-BN_APL2013}  Brillouin oscillation is caused by light interference between reflected light at the surface and diffracted light by the propagating acoustic pulse in the dielectric material, and its oscillation period corresponds to half of the wavelength of probe light in the material. However, it is difficult to accurately measure the elastic constant of dielectric thin films thinner than $\sim$100 nm because we can not determine Brillouin-oscillation frequency in a thinner film than the wavelength of probe light in it.

In this study, we propose a resonance method to measure the elastic constants of dielectric thin films thinner than $\sim$100 nm sandwiched by metallic thin films using picosecond ultrasonics. We make Pt/NiO/Pt three-layer films and excite the through-thickness resonances. In this structure, the strain energy of the first-resonance mode becomes larger in the NiO layer, which makes its frequency sensitive to the elastic constant of NiO ($C_{\rm NiO}$). Therefore, by measuring the first-mode resonant frequency, we can determine $C_{\rm NiO}$ even if the elastic constant of Pt ($C_{\rm Pt}$) includes uncertainty. Thinner Pt-film thickness makes the resonance frequency more insensitive to $C_{\rm Pt}$. However, we need the Pt layer to excite and detect the resonance. Therefore, we calculate the contribution of $C_{\rm NiO}$ to the resonance frequencies and propose an efficiency function by considering the contribution and excitation-detection efficiency of light to estimate the optimal Pt thickness in this structure. We prepare four structures by changing the Pt-film thickness remaining the NiO-film thickness. We observe 20--50 GHz resonances and compare the vibration amplitudes with the calculated efficiency function, and, we determine the optimal thickness to be $\sim$15 nm. We then make Pt/NiO/Pt films at different deposition temperatures from 300 to 700 K, and inversely determine $C_{\rm NiO}$. We also make Pt monolayers under the same conditions and measure their elastic constants to evaluate their effects on $C_{\rm NiO}$ in the inverse calculation. We discuss the effects of uncertainty in $C_{\rm Pt}$ and thickness of each layer, confirming the validity of this method.

\section{ Experimental methods}
\subsection{Thin films}
We used three substrates in this study: (0001)-Al$_2$O$_3$ (AO), (100)-Si, and (100)-Si with thermally oxidized surfaces (TO-Si). They were cleaned by ultrasonication in N-Methyl-2-pyrrolidone, isopropyl alcohol, and then distilled water. We further remove contamination on the surface by ambient-gas plasma cleaning for 10 min. We deposited Pt and NiO films by the RF magnetron sputtering method. Base pressure, Ar pressure, and sputtering power were $\sim$1$\times10^{-5}$ Pa, 0.8 Pa, and 50 W, respectively. We deposited thin films at three temperatures of 293, 553, and 706 K. We measured thickness and crystal structure by the X-ray reflectivity (XRR) and X-ray diffraction (XRD) methods with Co-K$\alpha$ X-ray, respectively.

\subsection{Picosecond ultrasonics}
We use a Ti-sapphire pulse laser whose wavelength and repetition rate are 800 nm and 80 MHz, respectively. We divide the pulse light into the pump and probe lights with powers of $\sim$20 mW and $\sim$5 mW, respectively. A corner reflector changes the arrival time of the pump light to the films. The pump light is modulated at 100 kHz by an acousto-optical crystal modulator. The probe-light wavelength is converted into 400 nm by a second harmonic generator. Both lights incident perpendicularly on the films through a 50-magnification objective lens, and the reflected lights are distinguished by a dichroic mirror, which reflects the 800-nm pump light and transmits the 400-nm probe light.  A part of probe light is collected by a balanced photodetector as a reference light, and the probe light reflected at film surface enters the balanced photodetector as well. Their subtraction signal is fed to a lock-in amplifier, to detect the reflectivity change due to the acoustic perturbation.

\section{Results and discussion}
\subsection{Efficiency function}
In this study, we made Pt/NiO/Pt films and determined the elastic constants of NiO from the multilayer resonance. We use a one-dimensional free-resonance model,\cite{QCM-ResoF, APEX-Uehara, NA-AlN_BAW} and calculate resonance frequencies $f$ and corresponding distributions of strain $\varepsilon(z)$ and strain energy $U(z)$ by solving the frequency equation, where $z$ denotes the coordinate along the thickness direction. Here, we briefly describe the procedure. Assume that the longitudinal plane waves $u_m = A_m e^{-i k_m z} + B_m e^{i k_m z}\, (m=1,2,3)$ propagate in each layer, where $u_m, k_m, A_m$, and $B_m$ denote displacement, wave vector, and amplitudes in the $m$-th layer, respectively. Considering the continuities of stress and displacement at the interfaces and stress-free conditions of the top and bottom surfaces, we obtain six equations. Using the $6 \times 6$ matrix $\Gamma$ and the amplitude vector $\bm{A} = (A_1, A_2, A_3, B_1, B_2, B_3)$, we have $\Gamma \bm{A} = \bm{0}$. By solving $\det |\Gamma| = 0$, we obtain the resonance frequencies and calculate strain distributions from the amplitude vector. Here, we define the lowest-frequency solution as the first mode and the next as the second mode. The resonance frequencies depend on the film thickness and the sound velocity of each layer. Thinner Pt makes the resonance frequency insensitive to the elastic constant of Pt. To quantitatively evaluate this effect, we define the contribution $\beta_M$ of elastic constant $C_M$ to the resonance frequency $f$ as follows:
\begin{align}
\beta_M=\left| 2\frac{ \partial f}{\partial C_M} \frac{C_M}{f} \right| \label{eq-Contri},
\end{align}
where $M$ denotes Pt or NiO. This contribution is normalized to take a value between 0 and 1 and the summation of $\beta_M$ for each layer becomes 1 because the square of a resonant frequency is approximately expressed by the linear combination of the all elastic constants\cite{Ogi-RUS}.
 Thinner Pt increases $\beta_{\rm NiO}$ and decreases $\beta_{\rm Pt}$, and allows determining the elastic constant of NiO accurately.

However, it is difficult to excite and detect the resonance for the thinner Pt film because absorption and strain in Pt layer decrease. Therefore, Pt has an optimal thickness to determine $C_{\rm NiO}$, and we introduce the efficiency function to estimate it by considering the excitation and detection efficiencies for the resonance and the contribution of $C_{\rm NiO}$ to its frequency ($\beta_{\rm NiO}$). First, we consider the excitation efficiency in Pt/NiO/Pt films. The pump light causes thermal expansion in Pt layers, which is proportional to the intensity of the pump light $|E_{800}(z)|^2$. This thermal stress excites resonances, whose stress distributions are proportional to the strain distributions $\varepsilon(z)$. Therefore, we propose the excitation efficiency as a correlation function between them as follows: $\int_{\rm Pt} \overline{\varepsilon}(z) |E_{800}(z)|^2 dz$, where the integral is calculated in the Pt layers because NiO does not absorb 800-nm light. The normalized strain $\overline{\varepsilon}$ is defined so that the integral of the strain energy $\frac{1}{2} \int C \overline{\varepsilon}^2 dz$ takes unity.
We calculate the electric field $E(z)$ in Pt/NiO/Pt multilayers using a multi-reflection model (details are described in the supplementary material).

\begin{figure}[t]
\begin{center}
\includegraphics{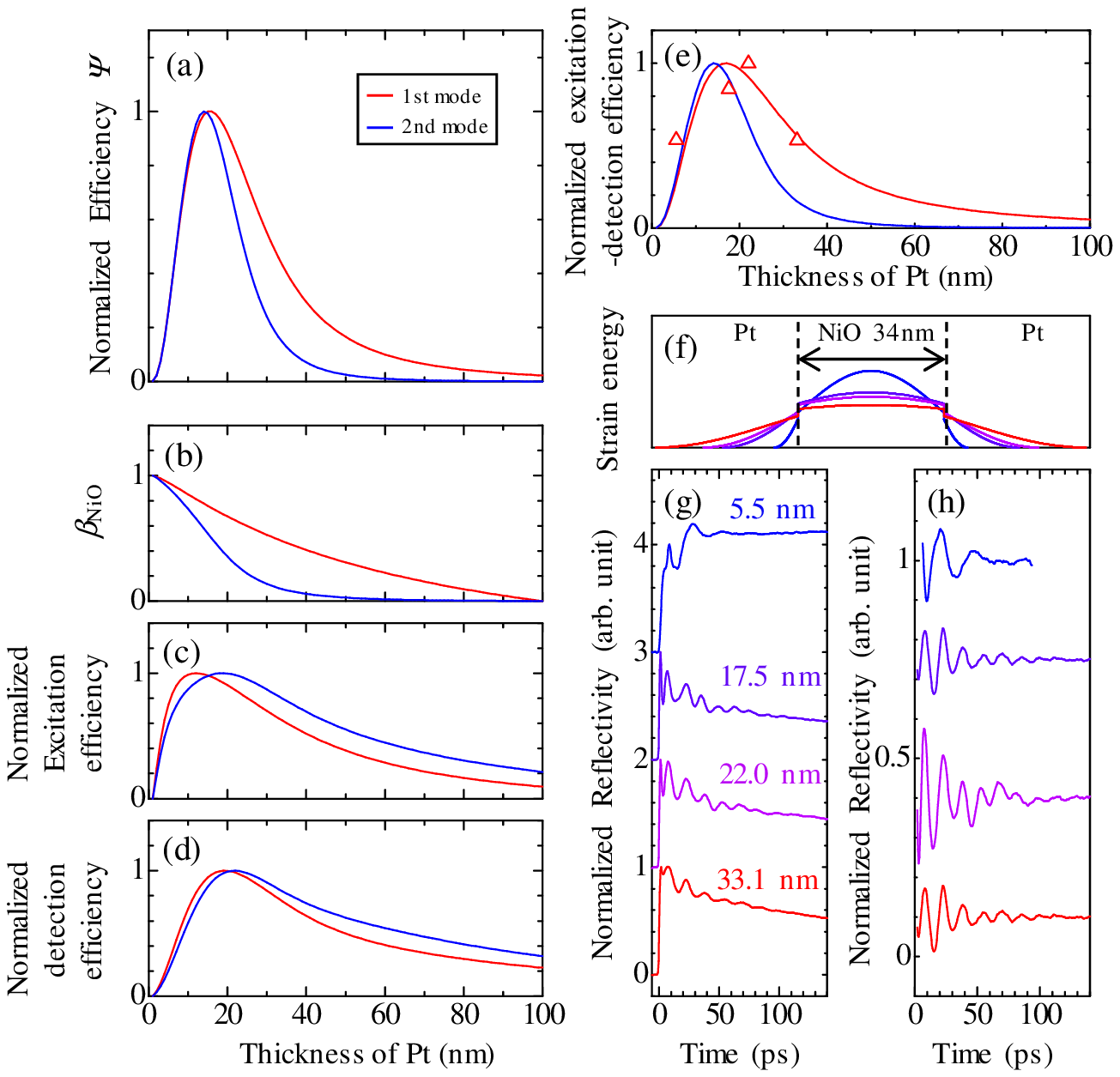}
\caption{}
\label{Efficiency}
\end{center}
\end{figure}

Next, we consider the detection efficiency. We observe the reflectivity changes from the photoelastic effect, which is proportional to the strain and intensity of the probe light. Therefore, we propose the detection efficiency as $\int_{\rm Pt} \overline{ \varepsilon }(z) |E_{400}(z)|^2 dz$, where the integral is calculated in the Pt layers because the probe light is mainly reflected by the Pt layers. Finally, the contribution of $C_{\rm NiO}$ to a resonance frequency is evaluated by $\beta_{\rm NiO}$ in Eq. (\ref{eq-Contri}). From the above discussion, we define the efficiency function $\Psi(d_{\rm Pt})$ to determine the optimal Pt thickness as follows:
\begin{align}
  \Psi (d_{\rm Pt}) = \beta_{\rm NiO} \int_{\rm Pt} \overline{\varepsilon}(z) |E_{800}(z)|^2 dz \times \int_{\rm Pt} \overline{\varepsilon}(z) |E_{400}(z)|^2 dz. \label{eq-Eff}
\end{align}
Figure \ref{Efficiency}(a) shows the calculated efficiency functions of the first and second modes for the multilayer system with NiO film with $d_{\rm NiO}=34$ nm. As the Pt thickness increases, the strain in Pt layers increases, leading to the increase in $\Psi$ until $d_{\rm Pt}\sim$15 nm. However, the effect of the decrease of $\beta_{{\rm NiO}}$ becomes more extensive, resulting in the peak of $\Psi$ around 13--18 nm. Figures \ref{Efficiency}(b)--(d) separately show the $\beta_{\rm NiO}$, the normalized excitation efficiency, and the normalized detection efficiency, respectively.

To verify this function, we prepared four Pt/NiO/Pt films and measured the resonance amplitude by changing only $d_{\rm Pt}$ between 5.5 and 33.1 nm. Figure \ref{Efficiency}(f) shows calculated strain-energy distributions of the first mode. Observed reflectivity changes are shown in Fig. \ref{Efficiency}(g), whose amplitudes are normalized by the intensity at initial absorption. We extract the resonance by subtracting the background change due to thermal diffusion using a polynomial function as shown in Fig. \ref{Efficiency}(h). The strain energy for the multilayer with 5.5-nm Pt films concentrates in the NiO layer, leading to a high $\beta_{\rm NiO}$. However, it is difficult to excite and detect its resonance as shown in Figs. \ref{Efficiency}(g) and (h). For multilayers with 17.5-nm and 22.0-nm Pt films clear resonances appear, and the resonance amplitude decreases in the multilayer with 33.1-nm Pt film. We integrate the resonance amplitudes and compare the values with the excitation-detection efficiency in Fig. \ref{Efficiency}(e). They agree with each other, validifying our model and usefulness of the proposed efficiency function $\Psi$. Thus, we conclude that the optimal Pt thickness is $\sim$15 nm for $d_{\rm NiO}=34$ nm.

\subsection{Measure the elastic constant of NiO}
To measure $C_{\rm NiO}$, we made Pt/NiO/Pt films and Pt monolayers on AO, Si, and TO-Si substrates by changing deposition temperature $T_d$ between 300 and 700 K. We measured $d_{\rm NiO}$ and $d_{\rm Pt}$ by the XRR method as listed in Table \ref{tabMV} (XRR spectra are shown in Fig. S2 in the supplementary material). XRD spectra shown in Fig. \ref{XRD} reveal that increase in $T_d$ enhances the crystallinity except $T_d=706$ K for Si substrate: At $T_d=293$ K, Pt shows (111) and (200) peaks, being polycrystalline. As $T_d$ increases, the intensities of Pt (111) and NiO (111) increase, and the Pt-(200) peak disappears. Interface roughness also decreases (Table S1). However, the intensities of Pt peaks become weak in the film deposited on Si substrate at $T_d=706$ K, and the film shows no NiO peaks and exhibits a Ni (111) peak, insisting that the crystal structure deteriorates and NiO decomposes.

We observe the first-resonance mode for all films and the second mode for the films deposited at $T_d=293$ K (Fig. S3). We measured the resonance frequencies at 4--9 different points for each film, and inversely determined $C_{\rm NiO}$ by minimizing $\sum_j \left(1-f_j^c/f_j^m \right)^2$, where $f_j^c$ and $f_j^m$ are calculated and measured resonance frequencies of the $j$-th mode. To calculate $f_j^c$, we need the value of $C_{\rm Pt}$, and we measured it for the Pt monolayers deposited under the same condition as listed in Table \ref{tabMV}. We measured the thickness of them by the XRR method (Table S2 and Fig. S4(a)) and observe $\sim$130 GHz resonances (Figs. S4(b-d)). We also use a bulk value $C_{{\rm Pt}}$=384.7 GPa\cite{EC-Pt_lowTemp} along $\braket{111}$ direction to determine $C^*_{{\rm NiO}}$.
We use the bulk values for mass density because calculated XRR spectra with the bulk densities well agree with measured spectra (Fig. S2). Even if the densities of NiO and Pt deviate from the bulk value by 1\%, for example, the determined elastic constant changes $\sim$1\%, and it is within the measurement error.

Determined $C_{\rm NiO}$ and $C_{\rm NiO}^*$ are shown in Fig. \ref{EC-NiO} and Table \ref{tabMV}. They agree with each other within a 3\% deviation, which is smaller than $\sim$7\% measurement deviations. $C_{\rm NiO}$ at $T_d=293$ K is smaller than the corresponding bulk value\cite{NiO-EC} by $\sim40$\%, and it increases with $T_d$, approaches to the bulk value except for the Si substrate. Thin films contain many defects, and annealing treatment decreases defects at grain boundaries, resulting in the recovery of elastic constants.\cite{Pu-Naka-CuAnneal, JJAP-Nakamihci} However, on the Si substrate, the increase in $T_d$ decomposes the crystal structure, decreasing the elastic constant. $C_{\rm NiO}$ of the films on AO substrate is smaller than that on TO-Si substrate by $\sim$16\% at each temperature, and we consider that difference in crystallinity and interference structure due to substrates causes this softening.

\subsection{Effects of other film parameters}
To evaluate the determination error due to uncertainty of $C_{\rm Pt}$ and film thickness, we calculate resonance frequencies of first and second modes using bulk values, and inversely determined $C_{\rm NiO}$ from them by changing $C_{\rm Pt}$, $d_{\rm Pt}$, or $d_{\rm NiO}$ within $\pm10$\% in the inverse calculation. The deviations $\delta C_{\rm NiO}$ from the initial value are shown in Fig. \ref{Error}. $C_{\rm Pt}$ of the three-layer structure might be different from that of the monolayer even if they are deposited under the same condition. $C_{\rm Pt}$ of monolayer also contains measurement deviations. However, 5\% changes of $C_{\rm Pt}$, for example, causes $\sim$3\% deviations in $C_{\rm NiO}$, which is smaller than typical measurement deviations of $\sim$5\%. On the other hand, 5\% change of film thickness causes $\sim$8\% deviations in $C_{\rm NiO}$. Therefore, the uncertainty of $C_{\rm Pt}$ less affects $C_{\rm NiO}$ in this structure. We measure film thicknesses by the XRR method, where measurement uncertainty is $\sim$1\% at most, leading to 2\% deviations in $C_{\rm NiO}$. Therefore, we conclude that this three-layer resonance method enables us to measure the elastic constant of $\sim$30 nm dielectric films within measurement deviations.

\section{Conclusions}
To measure the longitudinal elastic constant of $\sim$30 nm dielectric films, we used a three-layer structure. We proposed an efficiency function to estimate the optimal thickness to determine the elastic constant of dielectric films by considering its contribution to the resonance frequency and excitation-detection efficiency. We prepared four films by changing Pt-layer thickness as 5.5, 17.5, 22.0, and 33.1 nm and measured their resonance amplitudes. Both calculated efficiency and measured amplitudes have a peak around 15 nm, validifying our model. We concluded that the optimal Pt thickness is 15 nm for Pt/NiO/Pt structure for 34-nm NiO films. We deposited the optimal-thickness Pt/NiO/Pt films by changing deposition temperatures between 300 and 700 K. We measured the resonance frequencies by picosecond ultrasonics, and inversely determined the elastic constant of NiO. The films deposited at room temperature have $\sim$40\% smaller elastic constant than a bulk. However, as the deposition temperature increases, crystallinity becomes better, and the elastic constant approaches to the bulk value. We evaluated the determination error of the elastic constant of NiO in this method, revealing that the uncertainty of the elastic constant of Pt is smaller than measurement deviations, and film thickness is more sensitive. We measured film thicknesses by the X-ray reflection method within $\sim$1\% uncertainty, leading to 2\% deviations in the elastic constant of NiO. We concluded that this method enables us to measure the elastic constant of dielectric thin films thinner than $\sim$100 nm accurately.

\acknowledgment
This work was supported by KAKENHI Grant No. 18H01859 of Grant-in-Aid for Scientific Research (B).

\clearpage
CAPTIONS:

\bigskip
Table 1:

Measured film thicknesses of Pt ($d_{\rm Pt}$) and NiO ($d_{\rm NiO}$) of Pt/NiO/Pt films, and the longitudinal elastic constant of Pt ($C_{\rm Pt}$) of monolayer films at each deposition temperature $T_d$. $C_{\rm NiO}$ and $C_{\rm NiO}^*$ are the inversely determined elastic constants of NiO using $C_{\rm Pt}$ of the monolayer films and a bulk, respectively. The standard deviations of $C_{\rm NiO}^*$ are equivalent to those of $C_{\rm NiO}$.

\bigskip
Figure 1

(a) Calculated efficiency function $\Psi$ for $d_{\rm NiO}=34$ nm, and its Pt-thickness dependence of each term: (b) the contribution, (c) the normalized excitation efficiency, and (d) the normalized detection efficiency. (e) The excitation-detection efficiency. Triangle plots represent the measured resonance amplitudes of Pt/NiO/Pt films. (f) Calculated strain-energy distributions for $d_{\rm Pt}=5.5$, 17.5, 22.9, and 33.1 nm. (g) Observed reflectivity changes of Pt/NiO/Pt films deposited on AO substrate at room temperature, whose intensity is normalized at initial absorption intensity, and (g) extracted resonances.

\bigskip
Figure 2

XRD spectra of Pt/NiO/Pt films deposited on AO, TO-Si, and Si substrates at different temperatures.

\bigskip
Figure 3

Determined longitudinal elastic constant $C_{\rm NiO}$ of Pt/NiO/Pt films of $d_{\rm NiO}$ $\sim$34 nm and $d_{\rm Pt}$ $\sim$18 nm. Closed symbols with solid lines and open symbols with dashed lines denote that we use the elastic constant of Pt of corresponding monolayer films and bulk values in the inverse calculation, respectively.

\bigskip
Figure 4

Evaluated determination error $\delta C_{\rm NiO}$ in the elastic constant of NiO due to the changes of elastic constant of Pt (dashed line), film thicknesses of Pt and NiO (solid lines).

\begin{table}[t]
  \centering
  \caption{}
    \begin{tabular}{ccccccc}
\hline \hline
    Substrate & $T_d$ & $d_{\rm Pt}$ & $d_{\rm NiO}$ & $C_{\rm Pt}$ & $C_{\rm NiO}$ &$C^{{*}}_{\rm NiO}$ \\
 & (K) & (nm) & (nm) & (GPa) &(GPa) &(GPa)\\
\hline
    AO & 293   & 17.52 & 34.42 & 369.9 $\pm$ 30.8 & 173.0 $\pm$ 12.5 & 171.6 \\
    & 553   & 17.77 & 26.05 & 362.2 $\pm$ 36.0 & 249.7 $\pm$ 21.4 & 245.2 \\
    & 706   & 17.94 & 27.24 & 377.6 $\pm$ 4.7 & 286.7 $\pm$ 44.1 & 284.8 \\
\hline
    TO-Si & 293   & 17.97 & 33.46 & 357.6 $\pm$ 8.5 & 204.3 $\pm$ 3.8 & 204.6 \\
    & 553   & 17.44 & 28.36 &       &       & 300.0 \\
    & 706   & 17.97 & 28.37 & 357.4 $\pm$ 2.8 & 341.3 $\pm$ 8.2 & 331.8 \\

\hline
    Si    & 293   & 17.67 & 33.70 & 368.4 $\pm$ 0.0 & 199.0 $\pm$ 3.7 & 196.4 \\
    & 553   & 17.11 & 26.18 &       &       & 232.6 \\
    & 706   &       &       &       &       & 119.4 \\
\hline \hline
    \end{tabular}%
  \label{tabMV}
\end{table}%

\clearpage

\clearpage

\begin{figure}[t]
\begin{center}
\includegraphics{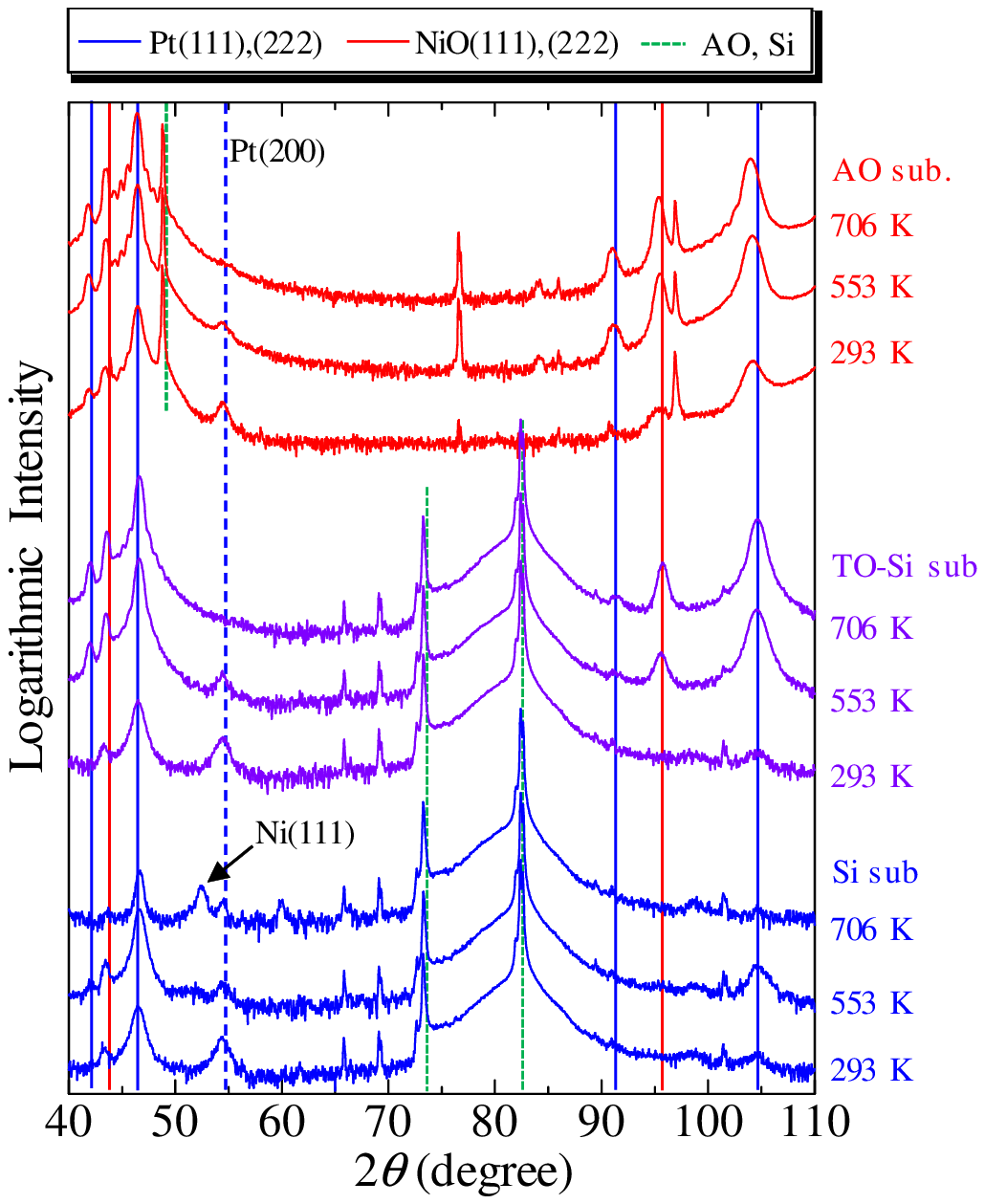}
\caption{}
\label{XRD}
\end{center}
\end{figure}

\clearpage

\begin{figure}[t]
\begin{center}
\includegraphics{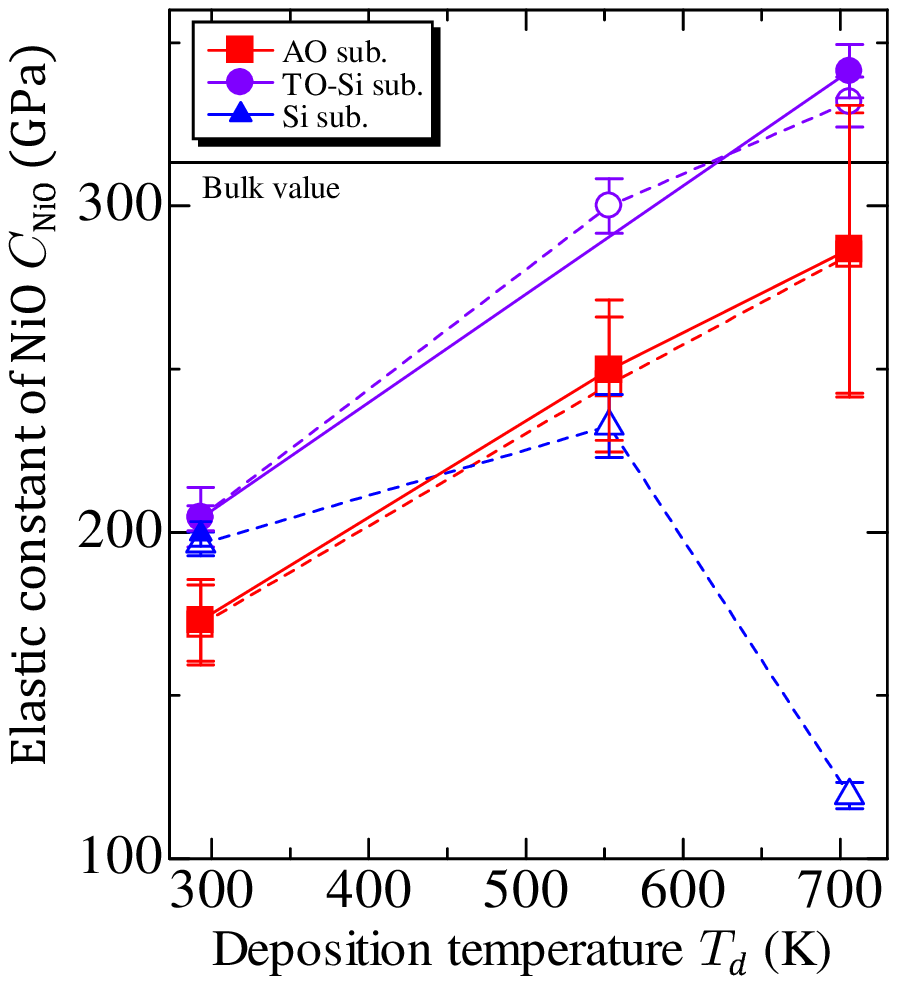}
\caption{}
\label{EC-NiO}
\end{center}
\end{figure}

\clearpage

\begin{figure}[t]
\begin{center}
\includegraphics{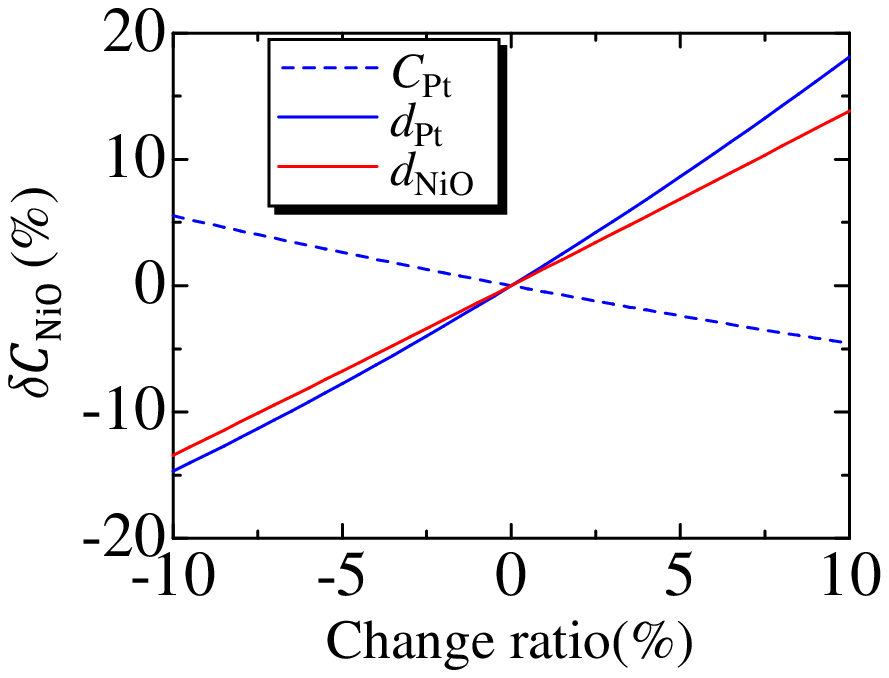}
\caption{}
\label{Error}
\end{center}
\end{figure}

\end{document}